\documentclass[useAMS,usenatbib]{mn2e}
\usepackage{epsfig}

\def\bs{\bigskip}

\def\ea{\ et al. \,}
\def\eg{{\it e.g.\,}}
\def\be{\begin{equation}}
\def\ee{\end{equation}}

\begin{document}

\title{CMB POLARIZATION DUE TO SCATTERING IN CLUSTERS}
\author[M. Shimon, Y. Rephaeli, B. W. O'Shea and  M. L. Norman]
{M. Shimon$^1,$$^2$, Y. Rephaeli$^1,$$^2$, B. W. O'Shea$^2,$$^3,$$^4$ and 
M. L. Norman$^2$\\
\\
\\
1.\ School of Physics and Astronomy, Tel Aviv University, Tel Aviv, 69978, Israel \\ 
2.\ Center for Astrophysics and Space Sciences, University
of California, San Diego, 9500 Gilman Drive, La Jolla, CA, 92093-0424\\ 
3.\ Department of Physics, University of Illinois at
Urbana-Champaign, 1110 West Green Street, Urbana, IL 61801-3080\\ 
4.\ Theoretical Astrophysics Group, Mail Stop B227, P.O. Box 1663, Los
Alamos National Laboratory, Los Alamos, NM 87545}
                                                                
\date{}

\maketitle

\begin{abstract}

Scattering of the cosmic microwave background (CMB) in clusters of 
galaxies polarizes the radiation. We explore several polarization 
components which have their origin in the kinematic quadrupole 
moments induced by the motion of the scattering electrons, either 
directed or random. Polarization levels and patterns are determined 
in a cluster simulated by the hydrodynamical Enzo code. We find 
that polarization signals can be as high as $\sim 1 \, \mu$K, a 
level that may be detectable by upcoming CMB experiments.

\end{abstract}

\begin{keywords}
cosmic microwave background, scattering, polarization.
\end{keywords}

\section{INTRODUCTION}
\label{INTRODUCTION}

Scattering of the CMB in clusters of galaxies imprints spectral changes 
on the Planck spectrum - the Sunyaev-Zeldovich (S-Z) effect - that 
constitute major cluster and cosmological probes. Cluster properties, 
such as intracluster (IC) gas density, temperature, gas mass, total 
mass, and peculiar velocities, as well as values of the basic 
cosmological parameters, can be determined from measurements of the 
effect in a (sufficiently) large sample of clusters (as reviewed by 
Rephaeli 1995, Birkinshaw 1999, and Carlstrom \ea 2002). While this 
wealth of astrophysical and cosmological information can be extracted 
largely from measurements of the {\it total} intensity change due to 
Compton scattering, the polarization state is also of considerable 
interest. Various polarization components have been identified in 
the original work of Sunyaev \& Zeldovich (1980), and further elaborated 
upon by Sazonov \& Sunyaev (1999). Measurements of polarization patterns 
in clusters may yield additional information on the dynamical state of 
the cluster, the cluster morphology, and its velocity component 
transverse to the line of sight (los).

Linear polarization is produced by Compton scattering when the incident 
radiation either has a quadrupole moment, or acquires it during the 
(first stage of the) scattering process. The level of S-Z polarization 
is low since it is proportional to the product of the quadrupole moment 
and the Compton optical depth, $\tau$. Cluster polarization signals are 
predicted to be comparable to or below $\sim 1 \, \mu$K. The detection 
of polarized CMB signals at this level seems now feasible, and indeed 
anticipated by several experiments (\eg Bowden \ea 2004). The projected 
resolution and sensitivities motivate investigation of the polarization 
levels and patterns that are expected in clusters; this is the basic 
objective of the work described here. 
Weak polarization components that are of interest, but ones that are 
not explored here, are produced when the radiation is scattered during 
aspherical collapse of protoclusters, 
scattering in a moving cluster in which the radiation develops anisotropy 
due to lensing (Gibilisco 1997), or polarization produced by IC magnetic 
fields (Ohno et al. 2003) and relativistic magnetized plasma (Cooray, 
Melchiorri \& Silk 2002).

The dynamical state of clusters undergoing mergers of subclusters 
cannot be adequately represented by simple analytical models. Since 
the polarization levels and patterns are quite sensitive to the 
non-uniform distributions of the gas and total mass, peculiar 
and internal velocities, and the gas temperature profile, a 
realistic characterization of the polarization properties can 
only be based on the results of hydrodynamic simulations. In the  
first stage of a simulation-based study, we present here results 
for the S-Z polarization properties in clusters simulated by the 
hydrodynamic Enzo code. We illustrate these for a simulated 
$1.4\times 10^{15}\, M_{\odot}$ cluster.

In the next section we summarize the cluster simulation work whose 
results are employed here. Section 3 briefly describes the various 
polarization components induced by Compton scattering in clusters. 
In section 4 we describe our simulation-based results, and end with 
a discussion in section 5.

\section{SIMULATED CLUSTERS}

In this paper we use the data obtained from Enzo, an Eulerian adaptive 
mesh cosmology code (O'Shea \ea 2004 and references therein). This 
code solves dark matter N-body dynamics with the particle mesh technique. 
The Poisson equation is solved using a combination of fast Fourier 
transforms on the periodic root grid and multigrid techniques on 
non-periodic subgrids. Euler's hydrodynamic equations are solved 
using a version of the piecewise parabolic method (PPM) algorithm of 
Woodward \& Colella (1984) which has been modified for high-Mach flows 
such as those found in galaxy cluster simulations (Bryan et al. 1995). 
Block-structured adaptive mesh refinement using the technique of Berger 
\& Colella (1989) is used to follow the evolution of objects of interest. 
Enzo allows arbitrary integer ratios of parent and child grid resolutions 
and mesh refinement based on a variety of criteria, including baryon and 
dark matter overdensity or slope, the existence of shocks, Jeans length 
resolution, and cell cooling time. The code is parallelized using the 
MPI message-passing library\footnote{http://www-unix.mcs.anl.gov/mpi/}. 
Additional physics packages are available (but not used for the 
simulations in this paper) and are described in O'Shea et al. (2004). 
The Enzo code is publicly available on the world wide 
web\footnote{http://cosmos.ucsd.edu/enzo/}.

The simulation was initialized as follows: In order to find a galaxy 
cluster of appropriate size we initialized a simulation using only dark 
matter in a volume of space 256 h$^{-1}$ Mpc on a side at $z=30$ (when 
density fluctuations are still linear on scales relevant here). The 
simulation used a 256$^3$ root grid with 256$^3$ dark matter particles 
($M_{DM} = 1.19 \times 10^{11}~M_{\odot}$). The standard $\Lambda$CDM 
cosmological model was 
used with $\Omega_{tot}=1$, $\Omega_{\Lambda}=0.7$, $\Omega_{M}=0.3$, 
$H_{0}=70$ km/sec/Mpc, $\sigma_{8}=0.9$, and $n=1$. Adaptive mesh 
refinement was turned on with an additional four levels of refinement 
(doubling the spatial resolution at each level) and the simulation 
was run to $z=0$.  At this point we stopped the simulation and used 
the Hop halo-finding algorithm (Eisenstein \& Hut 1998) to find the 
most massive dark matter halo in the simulation. This halo has a mass 
of $1.4 \times 10^{15}~M_{\odot}$ at $z=0$.

The simulation was then re-initialized at $z=30$ with both dark matter 
and baryons ($\Omega_{b} = 0.04,\, \Omega_{CDM} = 0.26$) with a 128$^3$ 
top grid and 2 static nested grids covering the Lagrangian volume where
the cluster forms, giving an effective root grid resolution of 512$^3$ 
cells (0.5 Mpc/h comoving) and dark matter particles of mass 
$1.49 \times 10^{10}$~M$_{\odot}$. Adaptive mesh refinement is allowed in 
the region where the galaxy cluster forms with a total of 7 levels of 
refinement beyond the root grid, for a maximum spatial resolution of 
15.625 h$^{-1}$~kpc (comoving). This simulation was then evolved to $z=0.06$, 
following the evolution of the dark matter and using adiabatic gas 
dynamics. The galaxy cluster of interest was then relocated with the Hop 
algorithm at $z=0.06$, $0.25$, $0.5$ and $1.0$, and data cubes with 
256$^3$ cells containing $\rho_{DM}$, $\rho_{gas}$, $T_{e}$ and the three 
baryon velocity components were extracted at these redshifts. The total 
cube length is 4 Mpc/h, corresponding to a spatial resolution of 15.625 
kpc/h (comoving). Unless otherwise noted, the results in the following 
sections are for the full 256$^3$ data cubes, and are based on the 
simulated cluster at $z=0.5$. The baryon density and velocity field are 
shown in Figure 1 and the cluster projected temperature and optical
depth are depicted in Figure 2.

\section{CLUSTER POLARIZATION COMPONENTS}

The main process that polarizes the CMB in clusters is Compton scattering 
of the radiation by IC electrons. Polarization levels depend on the cluster 
optical depth to Compton scattering, and either 
on the cluster velocity or gas temperature. 
These are typically in the $10-100$ nK range, but may reach values 
as high as $\sim \mu$K. We describe the main polarization components, 
and illustrate their predicted patterns in our fiducial
cluster. As we have noted, the CMB can be polarized by Compton 
scattering if the radiation multipole expansion has a finite 
quadrupole component in the electron rest frame. This can be 
brought about in several ways, as described in the following 
subsections. The level of linear polarization and its orientation 
are determined by the two Stokes parameters
\begin{eqnarray}
Q&=&\frac{3\sigma_{T}}{16\pi}\int n_{e} dl \int \sin^{2}\alpha\cos
2\psi I(\alpha,\psi) d\Omega \nonumber\\
U&=&\frac{3\sigma_{T}}{16\pi}\int n_{e} dl\int \sin^{2}\alpha\sin
2\psi I(\alpha,\psi) d\Omega ,
\end{eqnarray} 
where $dl$ is a length element along the photon path, $n_{e}$ is
the electron number density, $\sigma_{T}$ is the Thomson cross 
section, and $\alpha$ and $\psi$ define the relative directions of 
the incoming and outgoing photons. The average electric field
describes the polarization plane with a direction given by
\begin{eqnarray}
\varphi=\frac{1}{2}\tan^{-1}\frac{U}{Q}.
\end{eqnarray}
When the incident radiation is expanded in spherical harmonics
\begin{eqnarray}
I(\alpha,\psi)=\sum_{l,m}I_{lm}Y_{lm}(\alpha,\psi),
\end{eqnarray}
it becomes evident (from the orthogonality conditions of the spherical
harmonics) that only the quadrupole moment terms proportional to 
$I_{2,2}$ and $I_{2,-2}$ contribute to $Q$ and $U$ in Equation (1).

In the following subsections we describe the polarization produced by 
Compton scattering when a quadrupole moment is induced by electron motions. 
These `kinematic' polarization components arise due to either bulk (peculiar) 
or thermal (random) motion of the electrons. Since we are mainly interested 
in a more realistic assessment of the various components based on the 
results of hydrodynamic simulation (rather than idealized analytic models), 
previously derived theoretical expressions are only briefly summarized here.

\subsection{POLARIZATION INDUCED BY THE GLOBAL CMB QUADRUPOLE}

The CMB global quadrupole moment originates in the Sachs-Wolfe (SW) 
effect at the surface of last scattering, and later in the evolving 
gravitational fields of density inhomogeneities - the integrated 
Sachs-Wolfe (ISW) effect. Scattering by IC electrons polarizes the 
radiation at a level that is (linearly) proportional to the rms of the 
primary quadrupole moment and the cluster optical depth to Compton 
scattering, $\tau$ (Sazonov \& Sunyaev 1999). Using the WMAP normalization 
of the CMB quadrupole moment (Bennett \ea 2003) the maximal polarized 
signal is expected to be $\simeq 2.6\tau \, \mu K$, and its all-sky 
average is $\sim 60\%$ of this value (Sazonov \& Sunyaev 1999). 
The polarization pattern has broad peaks extending over an appreciable
fraction of the sky. The minima are narrow and correspond to the 
change of sign of the Stokes parameters. It has been noted that the 
dependence of this polarization component on the CMB quadrupole moment 
could possibly be exploited as a means to reduce cosmic variance 
(Kamionkowski \& Loeb 1997), and as a probe of dark energy models 
through the redshift evolution of the quadrupole 
(see, however, Portsmouth 2004). We note that the polarization due 
to the CMB quadrupole can be distinguished from the other cluster 
polarization components by virtue of its large-scale distribution 
reflecting that of the primary CMB quadrupole (\eg Baumann \& Cooray 
2003), and the fact that it is independent of frequency (when 
expressed in temperature units). This polarization component, which 
was recently studied in the context of N-body simulations by Amblard 
\& White (2004), will not be discussed further here.

\subsection{POLARIZATION DUE TO INTRINSIC QUADRUPOLE MOMENTS}

Scattering of the CMB in a cluster results in local anisotropy due to 
the different pathlengths of photons arriving from various directions to 
a given point. The optical depth effectively develops a dipole moment, 
and $\Delta T/T$ would appear anisotropic, except in the center of a 
spherically symmetric cluster, where electrons effectively 'see' almost 
isotropic (singly) scattered radiation. The local anisotropy of 
$\Delta T/T$ resulting from single scatterings is due to either 
thermal or bulk electron motions, constituting the well known thermal 
and kinematic components of the S-Z effect. This anisotropy provides 
the requisite quadrupole moment; second scatterings then polarize 
the radiation.

Two polarization components are directly induced by scattering in a 
moving cluster. The first is linear in the cluster velocity 
component transverse to the los, $v_t \equiv c\beta_{t}$, but quadratic in 
$\tau$; the second is linear in $\tau$ but quadratic in $\beta_{t}$.
A third polarization component arises from second scatterings (and, 
therefore, is $\propto \tau^2$) after local anisotropy is produced by the 
thermal S-Z effect. The spatial patterns of the various polarization 
components can be readily determined only when the morphology of the 
cluster and its IC gas are isotropic. The polarization patterns arising 
from scattering off thermal electrons are isotropic (Sazonov \& Sunyaev 1999) 
in a spherical cluster, while the corresponding patterns of the kinematic 
components are clearly anisotropic due to the asymmetry introduced by the 
direction of the cluster motion. 

To calculate the polarization components, we first note that the 
expressions for the $Q$ and $U$ Stokes parameters are specified 
in Equation (1) in terms of the relative directions of the ingoing 
and outgoing photons. Due to the obvious dependence of the angles and 
frequencies on the relative directions of the electron and photon 
motions, the expressions for $Q$ and $U$ are transformed to a frame 
whose $Z$ axis coincides with the direction of the electron velocity, 
the axis with respect to which the incoming and outgoing photon 
directions are defined (in accord with the choice made by Chandrasekhar 
1950). Since the Doppler effect 
depends only on the angle between the photon wave vector and the 
electron velocity, the calculation of the temperature anisotropy and 
polarization is easier 
if we assume the incident radiation is isotropic in the CMB frame 
(which is indeed the case to the level of accuracy relevant to this 
work). With the polarized cross section for Compton scattering (in the 
Thomson limit) written in terms of angles measured with respect to 
the electron velocity (after averaging over the azimuthal angles),
\begin{eqnarray}
\frac{d\sigma}{d\Omega}=\frac{3\sigma_{T}}{8}
\left(1-\mu_{0}^{2}\right)P_{2}(\mu'_{0})\, ,
\end{eqnarray}
where $P_{2}(\mu'_{0})$ is the second Legendre polynomial, the 
expression for the $Q$ parameter of the scattered radiation is 
\begin{eqnarray}
Q(\mu)=\frac{3}{8}\tau(1-\mu_{0}^{2})
\int_{-1}^{1}P_{2}(\mu'_{0})I(\mu'_{0})d\mu'_{0}\, .
\end{eqnarray}
Here, $\mu'_{0}=\cos\theta'_{0}$, $\mu_{0}=\cos\theta_{0}$, are the 
angle cosines between the electron velocity and the incoming and 
outgoing photons, respectively.

\subsubsection{The $\tau^{2}\beta_{t}$ and $\tau \beta_{t}^{2}$ Components}

The degree of polarization induced by the kinematic S-Z component was 
determined by Sunyaev \& Zeldovich (1980) in the simple case of uniform 
gas density,
\begin{eqnarray}
Q=\frac{1}{40}\frac{x e^{x}}{e^{x}-1} \tau^{2} \beta_{t} ,
\end{eqnarray}
where $x=h\nu/kT$ is the photon dimensionless frequency, with $T$ 
denoting the incident CMB temperature. 
A more complete calculation of this and the other polarization components 
was formulated by Sazonov \& Sunyaev (1999). Viewed along a direction 
$\hat{n}=(\theta,\phi)$, the temperature anisotropy at a point $(X,Y,Z)$, 
$\Delta T(X,Y,Z,\theta,\phi)$, generates polarization upon second 
scatterings. The Stokes parameters are calculated from Equation (1), 
\begin{eqnarray}
Q(X,Y)&=&\frac{3\sigma_{T}}{16\pi I_{0}}\int dZ n_{e}(X,Y,Z)\nonumber\\
&&\times\int d\Omega\sin^{2}(\theta)\cos(2\phi)\Delta I(X,Y,Z,\theta,\phi)\nonumber\\
U(X,Y)&=&\frac{3\sigma_{T}}{16\pi I_{0}}\int dZ n_{e}(X,Y,Z)\nonumber\\
&&\times\int d\Omega\sin^{2}(\theta)\sin(2\phi)\Delta I(X,Y,Z,\theta,\phi)
\end{eqnarray}
where now 
the $Z$ direction is chosen along the los, and $I_{0}$ is the incident 
intensity. The intensity change resulting from first scatterings is 
\begin{eqnarray}
& &\Delta I(X,Y,Z,\theta,\phi)/I_{0}=\sigma_{T}
\int d\vec{l}(X',Y',Z',\theta,\phi)\nonumber\\
& &\times n_{e}(X',Y',Z',\theta,\phi)\hat{n}\cdot 
\beta(X',Y',Z')\nonumber\\
& &\times\frac{x e^{x}}{e^{x}-1},
\end{eqnarray}
and the optical depth thru the point $(X,Y,Z)$ in the direction 
$(\theta,\phi)$ is
\begin{eqnarray}
\tau(X,Y,Z,\theta,\phi)=\sigma_{T}\int n_{e}(X',Y',Z')d\vec{l}
(X',Y',Z',\theta,\phi).
\end{eqnarray}
$Q(X,Y)$ and $U(X,Y)$ fully describe the 2-D polarization field.

The second kinematic polarization component is 
$\propto \tau \beta_{t}^{2}$; this component is generated by virtue of 
the fact that the radiation appears anisotropic in the electron frame. 
Expanding the apparent angular distribution of the radiation in Legendre 
polynomials, and keeping terms up to $\beta^2$, the quadrupole moment is 
determined. The polarization of the singly scattered radiation is then 
calculated (Sazonov \& Sunyaev 1999) by using 
Equation (5). 
(Note that 
the transformation back to the CMB frame does not change the polarization 
since this introduces additional terms which are higher order in $\beta$.) 
The level of this polarization component (which was originally determined 
by Sunyaev \& Zeldovich 1980) is 
\begin{eqnarray}
Q=\frac{x^{2}e^{x}(e^{x}+1)}{20(e^{x}-1)^{2}}
\tau\beta_{t}^{2} .
\end{eqnarray}
In our chosen frame of reference the Stokes parameter $U$ vanishes due 
to azimuthal symmetry, so the total polarization amplitude, $P$, is 
equal to $Q$ and the polarization is orthogonal to $\beta_{t}$. 
Relativistic corrections (to the non-relativistic expression of Sunyaev 
\& Zeldovich 1980) were calculated by Challinor, Ford \& Lasenby (2000) 
and Itoh et al. (2000). These corrections generally amount to a 
$\sim 10\%$ reduction in Q. In our calculations we use their analytic 
approximation (to the exact relativistic calculation)
\begin{eqnarray}
Q&=&\frac{x^{2}e^{x}(e^{x}+1)}{10(e^{x}-1)^{2}}
\tau\beta_{t}^{2}\nonumber\\
& &\times\left[\frac{1}{2}F
+\Theta\left(3F-2\left(2F^{2}+G^{2}\right)+\frac{1}{2}F\left(F^{2}+2G^{2}
\right)\right)\right.\nonumber\\
&+&\left.\beta_{r}\left(-\frac{1}{2}F
+\frac{1}{4}\left(2F^{2}+G^{2}\right)\right)\right]
\end{eqnarray}
where $F=x\coth\frac{x}{2}$, $G=x/\sinh\frac{x}{2}$,
$\Theta$ is the gas dimensionless temperature, and $c\beta_{r}$ is the
radial component of the cluster velocity.

\subsection{POLARIZATION DUE TO SCATTERING BY THERMAL ELECTRONS}

Analogously to the $\tau^{2}\beta$ component discussed above, double
scattering off electrons moving with random thermal velocities can 
induce 
polarization whose level is proportional to $\tau^{2}\Theta$. The 
anisotropy introduced by single scatterings is the usual thermal 
component of the S-Z effect with intensity change $\Delta I_{t}$. 
We use an analytic approximation to the exact 
relativistic calculation for $\Delta I_{t}$ that was obtained by 
Itoh, Kohyama \& Nozawa (1998), and Shimon \& Rephaeli (2004).   
Using the expression for $\Delta I_{t}$ (from the latter paper) in 
Equation (7), the polarization can be calculated as discussed above 
\begin{eqnarray}
\Delta I_{t}(X,Y,Z,\theta,\phi)
=\frac{\sigma_{T}I_{0}x^{3}}{e^{x}-1}
\sum_{i=1}^{5}f_{i}(x)\int n_e({\bf r})\Theta({\bf r})^{i}dl. 
\end{eqnarray}
In Equation (12) 
the integration is along the photon path prior to the
second scattering, $f_{i}(x)$ are spectral functions of 
$x$ that are specified by Shimon \& Rephaeli (2004).

\section{RESULTS}

We have calculated the polarization patterns of the kinematic and 
thermal components using a simulation of a massive cluster 
at $z=0.50$, with $M_{cl} = 1.4\times 10^{15}\, M_{\odot}$ and 
velocity dispersion of $\approx 1000$ km/s.
The baryon density and the velocity field in the relevant region of the 
simulated cluster are shown in Figure 1. Polarization levels in all the 
figures are specified in terms of the equivalent (absolute) temperature. 
The figures show results in the sky plane with the los along the Z-axis.

The calculations of the double scattering components (proportional 
to $\tau^{2}\Theta$ and $\tau^{2}\beta$) are more challenging than 
that of the single scattering case. The calculations generally require 
ray tracing of doubly scattered photons. In Cartesian coordinates, 
Equations (7) \& (12) 
can be written as 
\begin{eqnarray}
Q(X,Y)&=&\frac{3\sigma_{T}^{2}}{16\pi}
\int\int dZd^{3}{\bf r'}n_{e}({\bf r})n_{e}({\bf r'})\nonumber\\
&&\times\frac{(X-X')^{2}-(Y-Y')^{2}}{|{\bf r'}-{\bf
    r}|^{4}}\sum_{i=1}^{5}f_{i}(x)\Theta({\bf r'})^{i}\nonumber\\
U(X,Y)&=&\frac{3\sigma_{T}^{2}}{8\pi}
\int\int dZd^{3}{\bf r'}n_{e}({\bf r})n_{e}({\bf r'})\nonumber\\
&&\times\frac{(X-X')(Y-Y')}{|{\bf r'}-{\bf r}|^{4}}
\sum_{i=1}^{5}f_{i}(x)\Theta({\bf r'})^{i}
\end{eqnarray}
where the first integration (over $Z$) is along the los. The electron 
density $n_{e}$ was calculated from the simulated $\rho_{g}$ taking a 
molecular weight of $\mu=0.63$.

Calculation of these 4-D integrals over the simulation $256^{3}$ data 
cells is computationally intensive, motivating an attempt to reduce the 
level of computation. For example, an approximation (employed by Lavaux 
et al. 2004) is to describe the first scattering by representing the 
gas in terms of a best-fit $\beta$ - profile. This significantly reduces 
the extent of the calculation. From Figures 1 \& 2, it is evident that 
our cluster cannot be fitted by a $\beta$ profile, isothermal or not; 
fitting 
merging clusters with subclumps and rich substructure with a $\beta$ 
profile is clearly inappropriate. Fortunately, Equation (13) 
can be 
cast in the form
\begin{eqnarray}
P_{+,\times}(X,Y)=\sigma_{T}^{2}\int dZ n_{e}({\bf r})h_{+,\times}({\bf r})
\end{eqnarray}
where $P_{+}$ and $P_{\times}$ are the Stokes $Q$ \& $U$ parameters,
respectively, and
\begin{eqnarray}
h_{+,\times}=f_{+,\times}\star g
\end{eqnarray}
is a convolution of the functions $f_{+,\times}$ with $g$, where
\begin{eqnarray}
f_{+}&=&\frac{X^{2}-Y^{2}}{r^{4}},\qquad\qquad f_{\times}=\frac{2XY}{r^{4}},\,\,\nonumber\\
g&=&\sum_{i}n_{e}({\bf r})f_{i}(x)\Theta({\bf r})^{i}.
\end{eqnarray}
This representation enables us to employ 3-D FFT calculations that are 
considerably faster than the direct integrations of Equations (13).
However, the full computation is still quite extensive; we therefore 
focus here on the core of the simulation volume.

Our results are summarized in Figure 3. Unlike the case of spherical 
(Sazonov \& Sunyaev 1999) or quasi-spherical (Lavaux et al. 2004) clusters, 
we obtain a more intricate 
polarization pattern and the quadrupole structure, which is absent in 
the spherically symmetric case, is evident. This pattern reflects the 
non-uniform gas distribution and an appreciable degree of non-sphericity 
of the cluster, which substantially affect primarily the 
double scattering polarization signals, essentially due to the fact that 
the second scattering is more likely to occur far from the cluster center 
(see also Sazonov \& Sunyaev 1999, Lavaux \ea 2004). The level of 
polarization at the 545 GHz Planck frequency band, at which the beam is 
4.5' FWHM, can reach $\sim 80$ nK, but when convolved with a narrow beam 
1' FWHM 
profile, its peak value can be as high as $\sim 0.7\mu K$. This level 
definitely grazes the threshold of next generation experiments. Note, 
however, that we have considered 
a rich cluster with maximal optical depth of $\tau_{0}\approx 0.01$ 
(Figure 2), but since this component is proportional to $\sim\tau^{2}$, 
typical values could be a factor $\sim 4$ lower or higher than the 
specific values quoted here. Additional uncertainty stems from the 
scaling of the cluster temperature with mass, and from the evolution of 
the temperature with redshift. Finally, this component was calculated 
at 545 GHz; measurements at lower frequencies will result in lower 
levels of polarization (Equation 13).

Equations (7)-(9)
for the double scattering kinematic component ($\propto 
\tau^{2}\beta$) can be written in the same form as 
Equation (14)
, but with
\begin{eqnarray}
h_{+,\times}=f_{+,\times}\star g\equiv\sum_{i=1}^{3}f_{i_{+,\times}}\star g_{i}
\end{eqnarray}
where
\begin{eqnarray}
f_{i+}=\frac{X_{i}(X^{2}-Y^{2})}{r^{5}} ,\,\,\,f_{i\times}=
\frac{2X_{i}XY}{r^{5}},\,\,\, g_{i}=n({\bf r})\beta_{i}({\bf r})
\end{eqnarray}
with $X_{i}$ ($i=1,2,3$) the components of the vector $(X,Y,Z)$. 
The results for the $\tau^{2}\beta$ component (for the central region 
of the simulation) are summarized in Figure 4. The level of polarization 
in absolute temperature units is $\sim 1-2$ orders of magnitudes smaller 
than the $\propto\tau^{2}\Theta$ component. Convolving with the Planck 
HFI and a gaussian beam profile with FWHM of 1', the peak values are 
$3$ nK and $25$nK, 
respectively, comparable to the level of the cosmological quadrupole 
polarization (section 3.1). As previously noted, both these components 
are frequency-independent, a fact that has obvious implications on the 
feasibility of separating these components out when the measurements 
will eventually reach the requisite sensitivity.

The polarization at 545 GHz due to the Doppler anisotropy, 
$\propto\tau\beta^{2}$, is plotted in Figure 5. At this high frequency 
the polarization
can be as high as $\sim 18$ nK 
when convolved with the Planck HFI. Convolution with a narrow gaussian 
with 1' FWHM beam yields higher ($\sim 0.1\mu K$) levels. Owing to its 
spectral dependence, this component can - in principle - be distinguished 
from both the $\propto\tau^{2}\Theta$ and $\propto\tau^{2}\beta$ polarization 
by multi-frequency measurements. Moreover, due to the different dependence 
on the optical depth, the two kinematic effects ($\propto\tau^{2}\beta$ 
and $\propto\tau\beta^{2}$) can possibly be separated out through their 
different morphologies (compare Figures 4 \& 5).

\section{DISCUSSION}

CMB observational capabilities have dramatically improved since 
the first measurement of the primary anisotropy by COBE/DMR. With 
projected sensitivities around 
$1$ $\mu$K, 
and sub-arcminute angular resolution, measurement of the dominant 
polarization signals in clusters by upcoming experiments seems feasible 
in the (not too distant) future.

The coupling of the CMB to IC gas by Compton scattering leads to various 
polarization components whose magnitudes and spatial patterns could 
provide significant information on the cluster motion and the dynamical 
state of IC gas. The feasibility of identifying a given polarization 
component (obviously) depends primarily on its magnitude, but this will 
be aided also by its particular spectral and spatial characteristics. 
The fact that some of the spatial patterns (such as that of the thermal 
S-Z polarization) produce no net signal (or substantially reduced signal 
in aspherical cluster) when 
integrated over a sufficiently large region of the cluster surface, 
can be exploited in optimizing the observational strategy for 
measurement of a specific polarization component.

In this work we have used cluster simulations to assess the amplitude
of three CMB polarization components. As expected, most 
polarization signals induced by the interaction of CMB photons and IC 
gas are substantially smaller than $1 \,\mu$K. An exception is the 
polarization induced by the double scattering thermal effect. For 
this component, convolving our simulation data with a FWHM 1' beam, 
we find that the effective polarization signal is about $\sim 1\mu K$. 
The polarization induced by double scattering off the thermal gas was 
the focus of the work by Lavaux \ea (2004). We have implemented a direct 
calculation scheme that yields results at the cluster core which 
qualitatively differ from those of the latter authors. By assuming 
isothermal $\beta$ profile to describe {\it first} scatterings, 
Lavaux \ea (2004) partially underestimated the effect of dynamics on 
producing the quadrupole in the first scattering; this may be a reason 
for the qualitative discrepancy between our respective results. By 
exploiting the properties of convolution integrals we are able to 
considerably facilitate the calculation of the components 
$\propto\tau^{2}\Theta$, $\propto\tau^{2}\beta$, essentially 
avoiding the need for 
simplifying approximations. The $\propto\tau^{2}\beta$ and 
$\propto\tau\beta^{2}$ components are an order of magnitude weaker, 
too small to be detected towards individual clusters in near future 
experiments.

This is the first stage of a comprehensive study of CMB polarization 
in clusters based 
on a realistic description of cluster structure that is afforded by 
hydrodynamic simulations. The results presented here already point 
to appreciable 
discrepancy with what is obtained when IC gas is modeled by a simple 
spherical distribution. We plan to extend the work presented here to a 
sample of simulated clusters with a range of masses and redshifts. The 
polarization signals in each of the sampled clusters together, 
supplemented by the expected confusing polarized signals, will then be 
analyzed in order to determine optimal strategies for the measurement 
of cluster-induced polarization components and the extraction of cluster 
properties.
\bs

\section*{Acknowledgments}

We thank the referee for several useful comments. Research at Tel Aviv 
University was supported by a grant from the Israel Science Foundation.
This work was supported in part by NASA grant NAG5-12140 and NSF
grant AST-0307690.  BWO has been funded in part under the auspices of
the U.S.\ Dept.\ of Energy, and supported by its contract W-7405-
ENG-36 to Los Alamos National Laboratory.  The simulations were
performed at SDSC with computing time provided by NRAC allocation
MCA98N020.

\newpage
\newpage
\newpage
 
\begin{figure}
\begin{center}
\epsfig{file=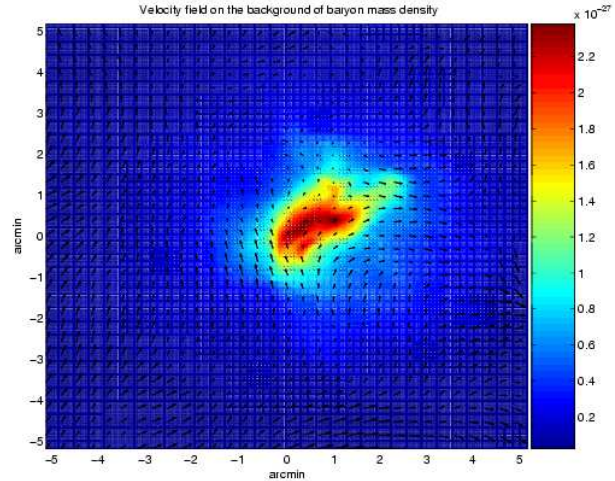,width=3.2in}
\end{center}
\caption{The velocity field is shown on the background of the baryon mass 
density; arrow length is linearly proportional to velocity magnitude. 
Color density scale is in units of $g/cm^{3}$.}
\end{figure}

\newpage

\begin{figure}
\begin{center}
\epsfig{file=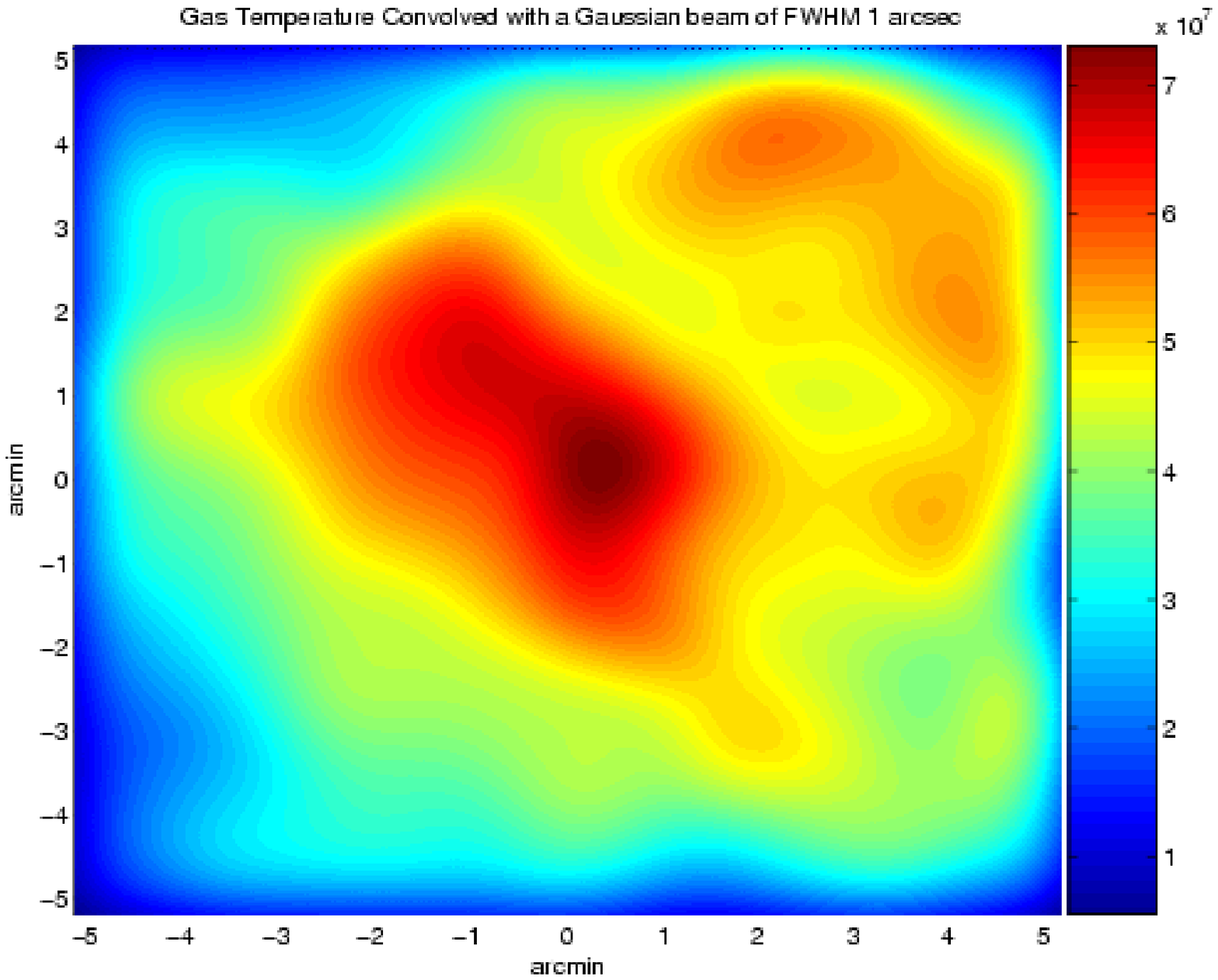,width=3.2in}
\epsfig{file=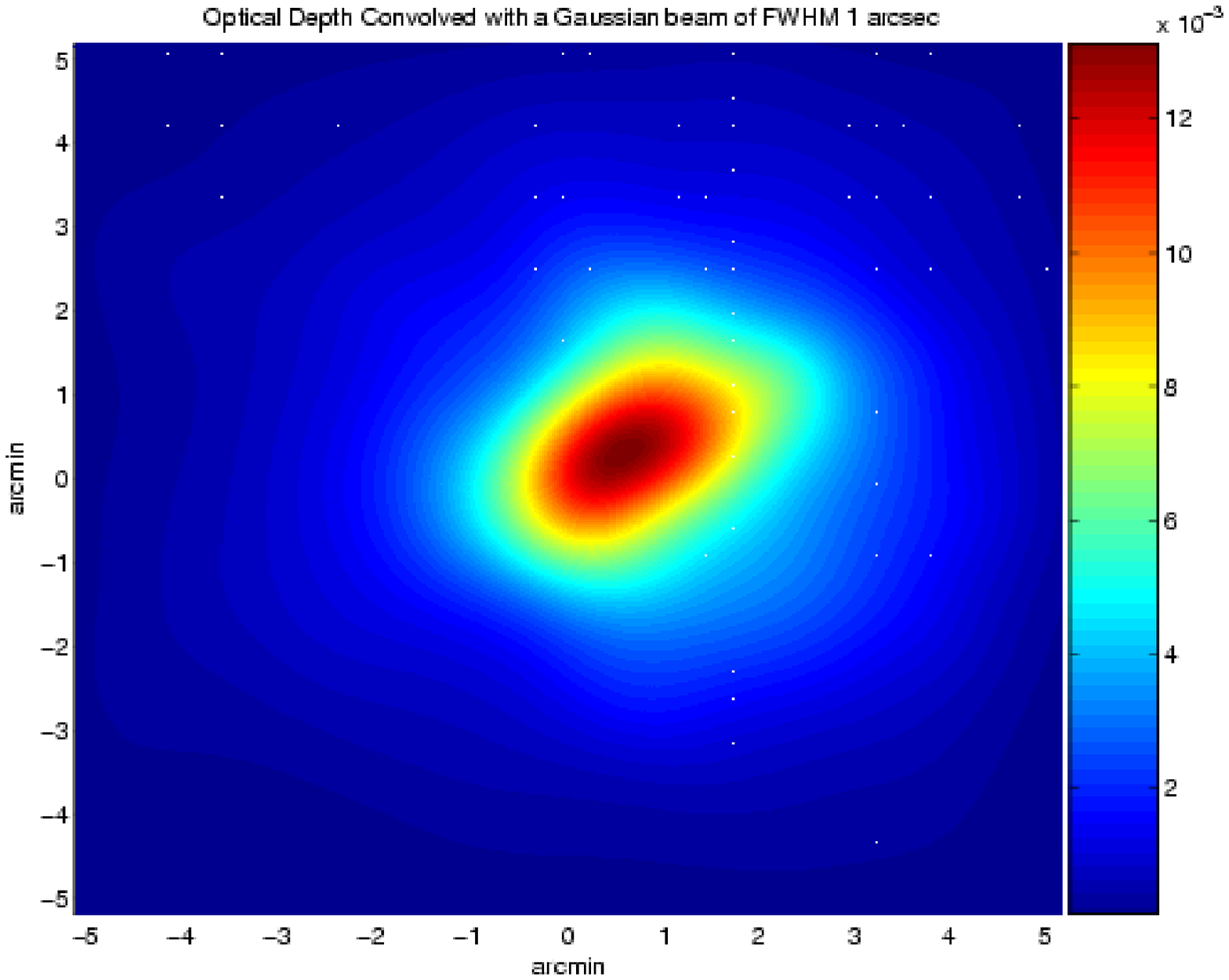,width=3.2in}
\end{center}
\caption{The gas temperature (top figure, with color scale in K) and optical 
depth (bottom)}
\end{figure}

\newpage

\begin{figure}
\begin{center}
\epsfig{file=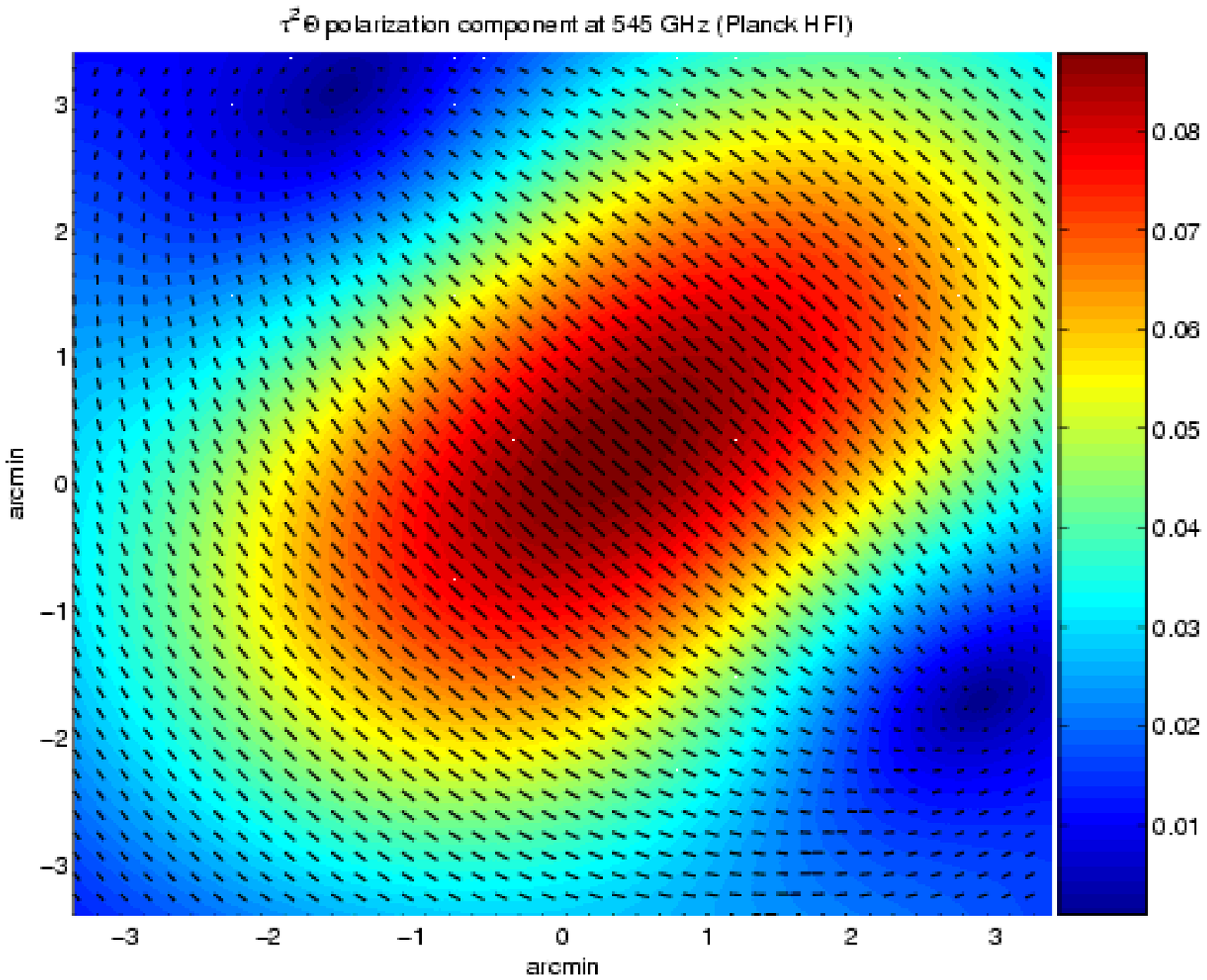,width=3.2in}
\epsfig{file=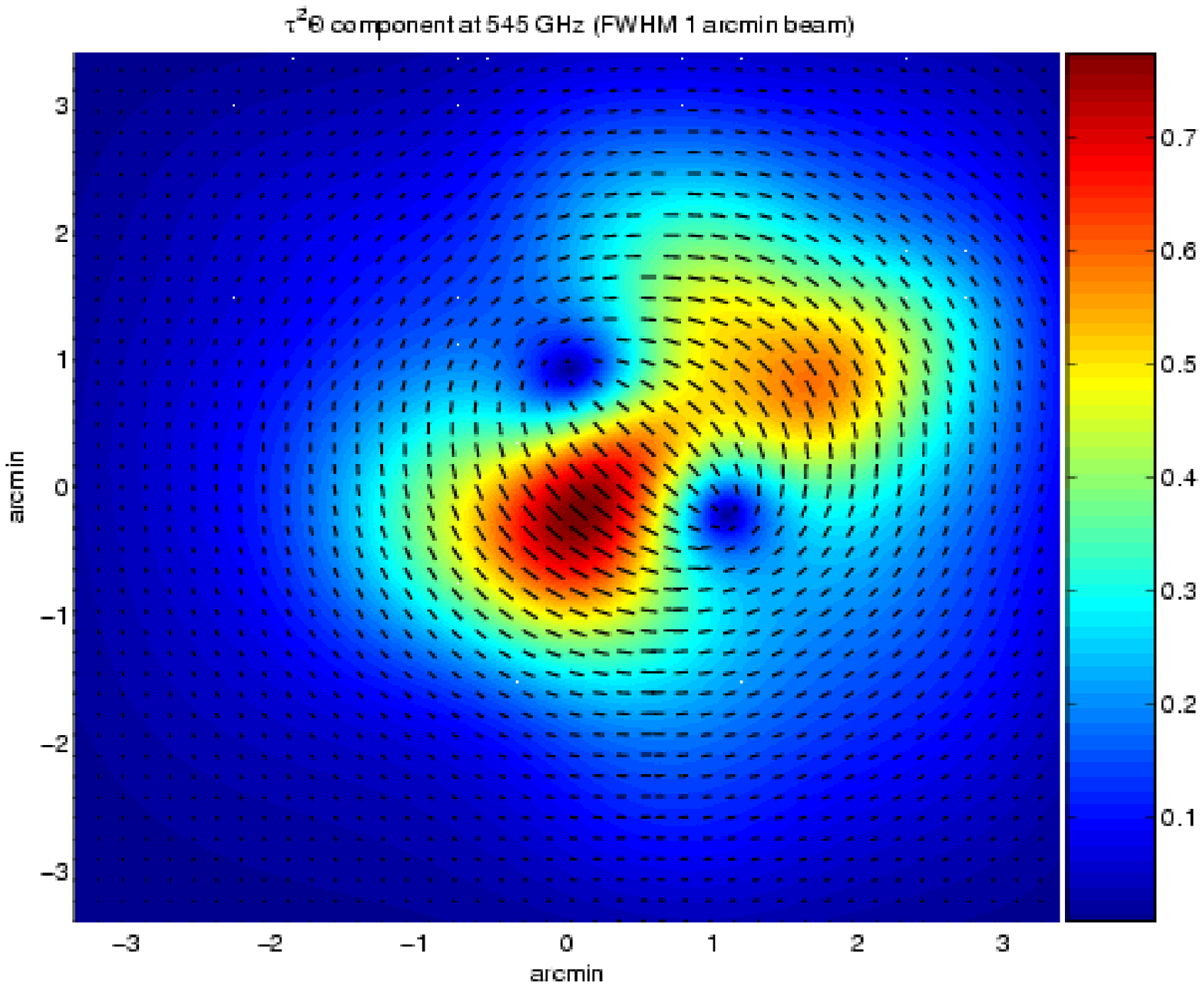,width=3.2in}
\end{center}
\caption{The $\tau^{2}\Theta$ polarization component convolved with the 
Planck HFI beam (top) and with a FWHM 1' beam (bottom). Color scale is 
in $\mu$K.}
\end{figure}

\newpage

\begin{figure}
\begin{center}
\epsfig{file=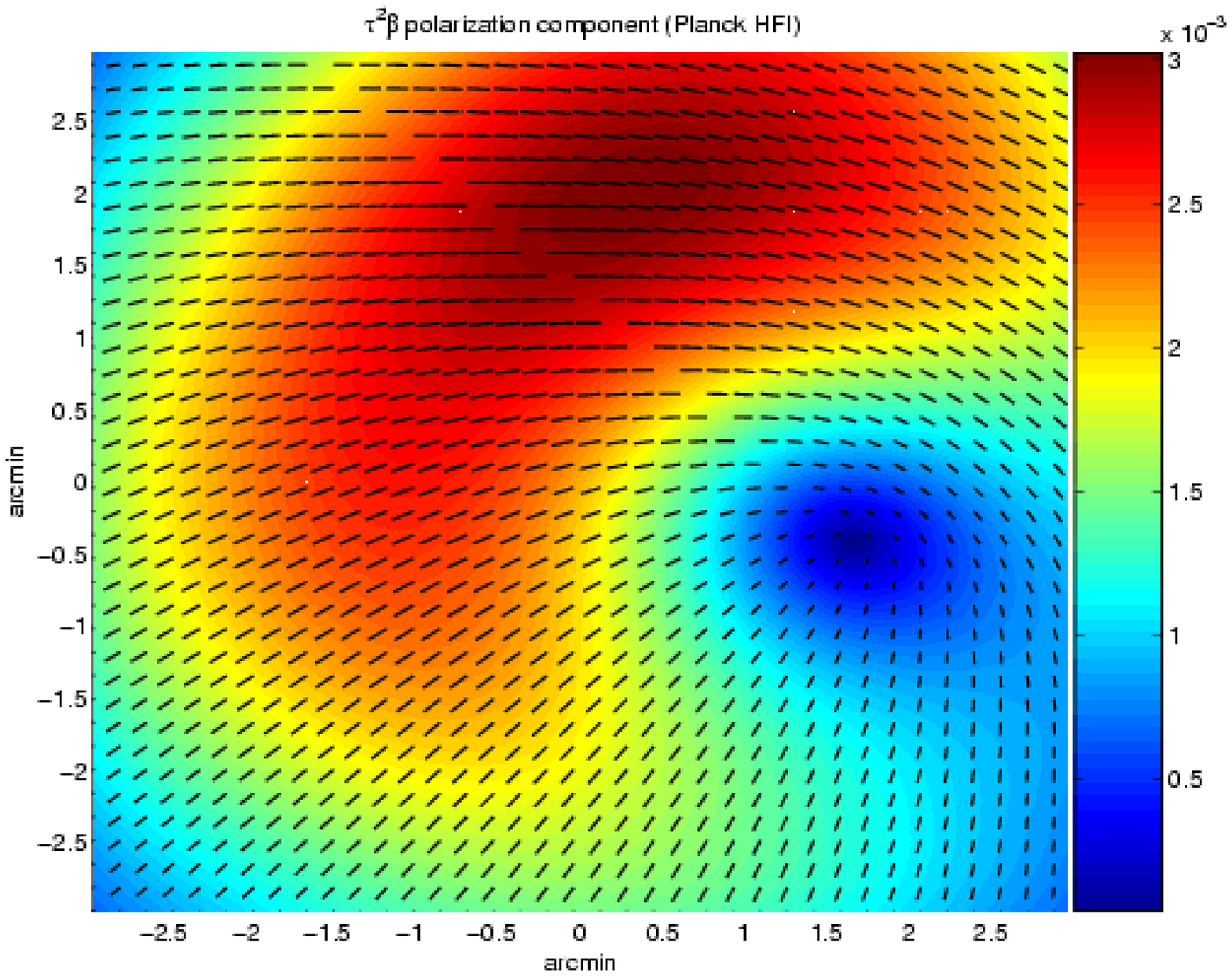,width=3.2in}
\epsfig{file=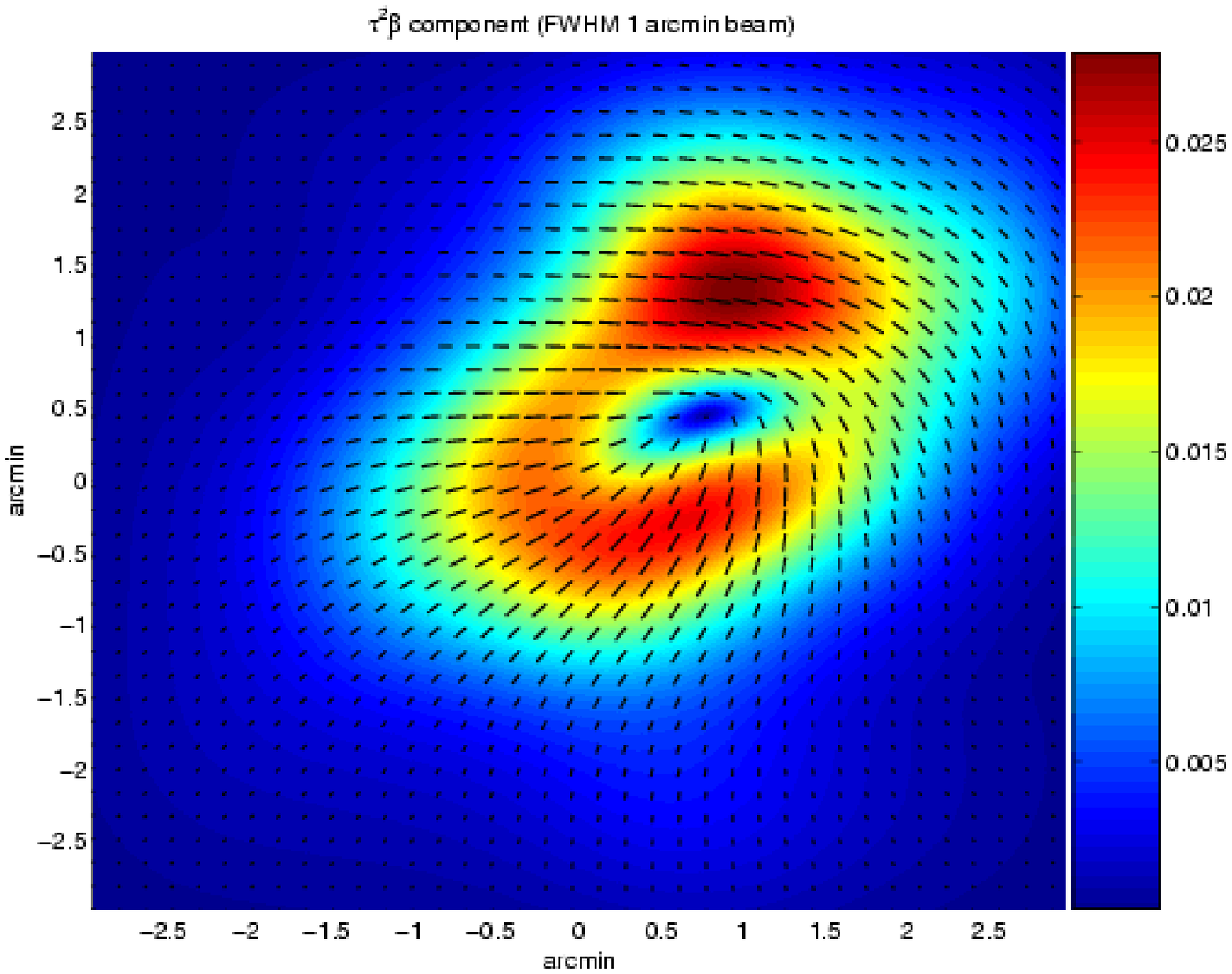,width=3.2in}
\end{center}
\caption{The $\tau^{2}\beta$ polarization component convolved with the 
Planck HFI beam (top) and with a FWHM 1' beam (bottom). Color scale is 
in $\mu$K.}
\end{figure}

\newpage

\begin{figure}
\begin{center}
\epsfig{file=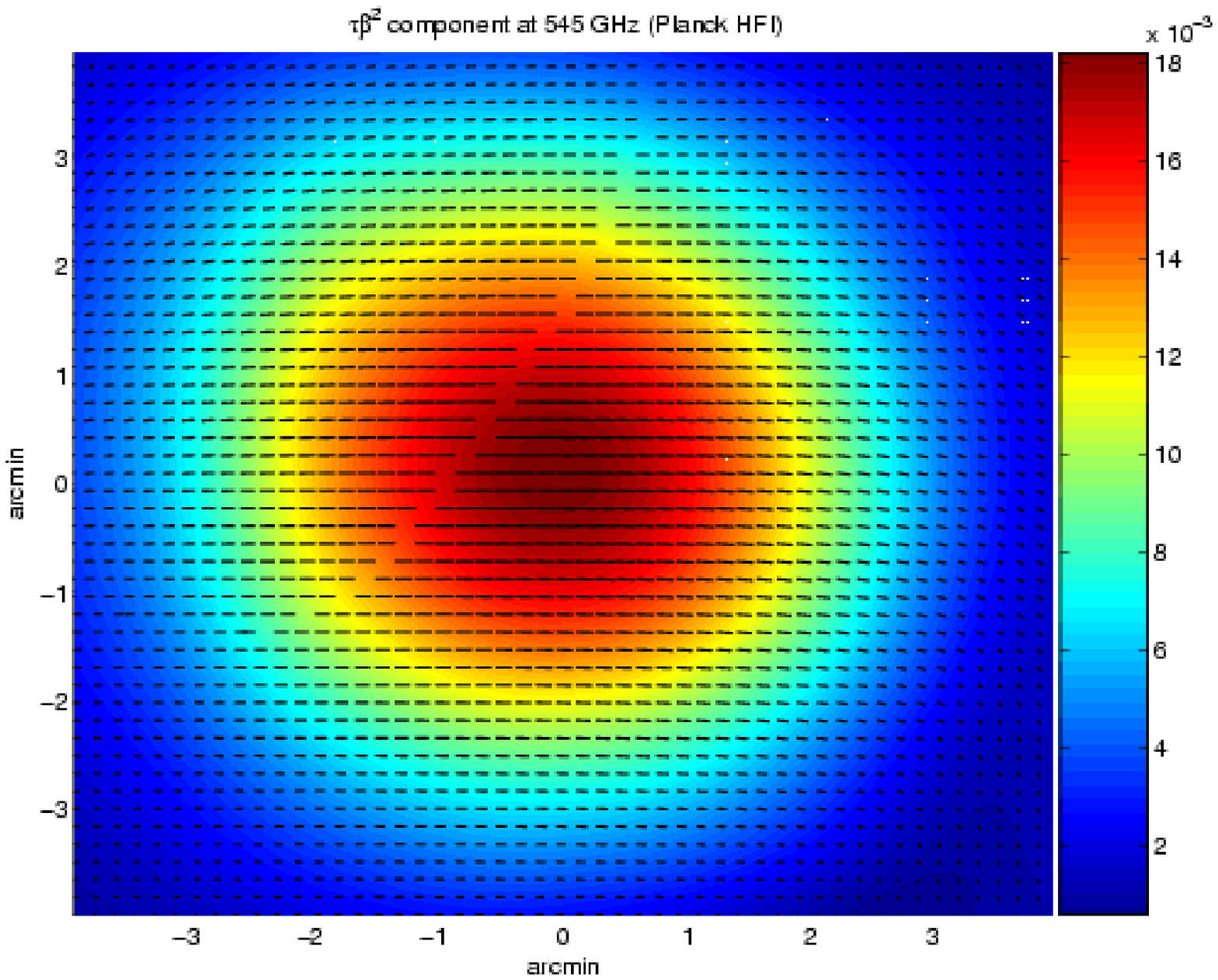,width=3.2in}  
\epsfig{file=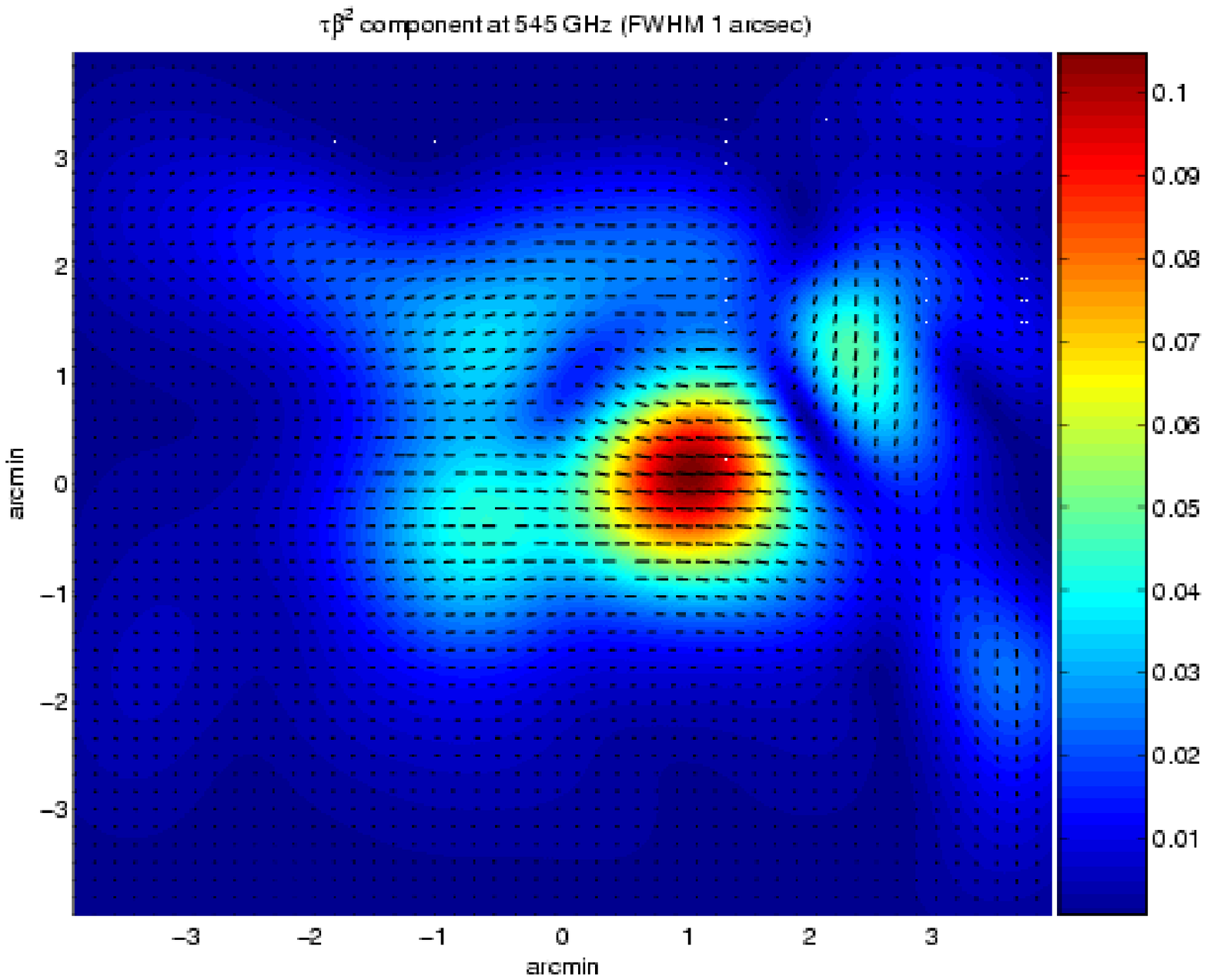,width=3.2in}
\end{center}
\caption{The $\tau\beta^{2}$ polarization component at 545 GHz 
convolved with the Planck HFI beam (top) and with a FWHM 1' beam 
(bottom). Color scale is in $\mu$K.} 
\end{figure}

\end{document}